\documentclass[twocolumn,amsmath,amssymb,showpacs,aps,prl]{revtex4}
\usepackage{graphicx}
\usepackage{bm}
\usepackage{epsfig}
\usepackage{color}
\newcommand \be{\begin{equation}}
\newcommand \ba{\begin{eqnarray}}
\newcommand \ea{\end{eqnarray}}
\newcommand \ee{\end{equation}}

\begin{document}


\title{Direct observation of L\'evy flight of holes in bulk $n$-InP}
\author{Serge Luryi }
\author{Oleg Semyonov}
\author{Arsen Subashiev}
\author{Zhichao Chen}
\affiliation{%
State University of New York at Stony Brook, Stony Brook, NY,
11794-2350}%

\begin{abstract}
We study the photoluminescence spectra excited at an edge side  of $n$-InP slabs and observed from the broadside. In a moderately doped sample the intensity drops off as a power-law function of the distance from the excitation --- up to several millimeters --- with no change in the spectral shape.The hole distribution is described by a stationary L\'evy-flight process over more than two orders of magnitude in both the distance and hole concentration. For heavily-doped samples, the power law is truncated by free-carrier absorption. Our experiments are near-perfectly described by the Biberman-Holstein transport equation with parameters found from independent optical experiments.
\end{abstract}
\pacs{05.40.Fb, 78.30.Fs,78.55.Cr,78.60.Lc,81.05.Ea}%
\maketitle

The transport of minority carriers produced by optical excitation in semiconductors 
is usually well described by a diffusion equation, characterized by a diffusion length $l^2=D\tau$, with $D$ being the diffusivity of carriers and $\tau$ their lifetime. The diffusion process can be viewed as the result of a random walk in which every step $x_i$ has the same probability density ${\cal P}(x_i)$. In moderately doped  direct-gap semiconductors, the transport of  minority  carriers is mediated by the transport of interband photons with carriers recombining and re-emerging repeatedly in a process called the ``photon recycling''. The random-walk steps $x_i$ are defined by the reabsorption length of photons produced in radiative recombination.  If the second moment of the distribution $\langle x_i^2 \rangle$ is finite, one has $D\propto \langle x_i^2 \rangle/\tau_i$, where $\tau_i$ is the mean time between steps (the radiative emission time $\tau_{\rm rad}$). This may lead to a substantial enhancement of $D$ and $l$ but the diffusion description would still be applicable. 

If, however, the step distribution ${\cal P}(x)$ is heavy-tailed, asymptotically satisfying a power law
\be
{\cal P}(x) \sim 1/x^{\gamma+1}
\label{StepDistr}
\ee
with $0 \le \gamma \le 1$, then the conventionally defined diffusivity diverges and the random walk is governed by rare but large steps \cite{Levy}.  Such a transport, called the L\'evy flight, features an anomalously large spread in space and a ``superdiffusive'' temporal evolution \cite{Bouchaud,Shlesinger,Metzler}.

Superdiffusive transport of light has  been known in the context of radiation trapping in propagation through media with narrow absorption lines.  This phenomenon occurs in different systems, ranging from stars \cite{Ivanov,Springmann} to dense atomic vapors \cite{Molisch}, such as gas lasers, discharges and hot plasmas. It has been studied for many decades,  starting from the theoretical papers by Biberman \cite{Bib} and Holstein \cite{Holstein}. Recently, L\'evy flights of photons were directly observed in engineered optical materials \cite{Wiersma} and in hot vapors \cite{Mercad}. Direct-gap semiconductors represent a new and exciting ``lab'' system for studying L\'evy flight of photons and/or minority carriers \cite{Semyon1,Luryi}. 
 
We have investigated \cite{Semyon1} the step distribution ${\cal P}(x)$ for the photon-assisted transport of holes in moderately doped $n$-InP and found that it asymptotically obeys the power law (\ref{StepDistr}) with $\gamma =0.7 \pm 0.1$. For heavier doping, the power law is truncated by free-carrier absorption,
\be
{\cal P}(x) \sim (1/x^{\gamma+1})~e^{-\alpha_{fc}x}~,
\label{StepDistrTrunc}
\ee
where $\alpha_{fc}$ [cm$^{-1}$] $\approx 1.3\times 10^{ -18} N_d$ is the free carrier absorption coefficient in InP \cite{AOS,Semyon2}.

The resultant random walk is limited by the loss of carriers in nonradiative recombination (of rate $\tau_{\rm nr}^{-1}$). The relative rates of recombination are characterized by the radiative efficiency $\eta = \tau_{\rm nr}/(\tau_{\rm nr}+\tau_{\rm rad})$ and the typical number of steps by the recycling factor $\Phi = \eta /(1- \eta )$. Due its high radiative efficiency (with $\Phi$ reaching $10^2$ for moderately doped samples), the $n$-doped InP is ideally suited for studying the L\'evy flight of holes \cite{AOS,Luryi,Semyon1,Semyon2}. The key parameters ($\gamma$ and $\Phi$) can be controlled by varying the doping $N_d$ and the temperature. The  truncation length $\alpha_{fc}^{-1}$ can also be controlled by varying $N_d$.

In earlier photoluminescence experiments with $n$-InP we found an indirect evidence of L\'evy flight of holes by analyzing the ratio of transmitted and reflected luminescence spectra across a thin flat wafer \cite{Semyon1}. 

\begin{figure}[t]
\begin{center}
\epsfig{figure=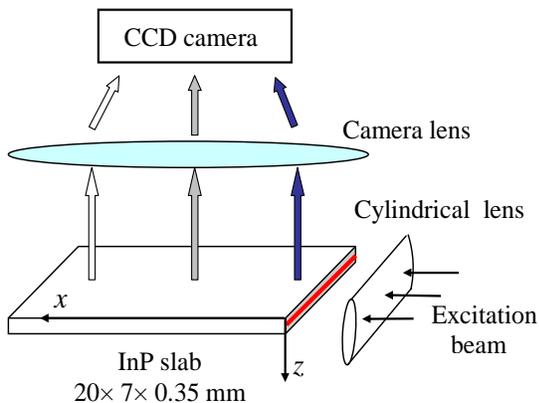,width=7.5cm,height=5.4cm} \caption[]
{\label{geom} Geometry of our photoluminescence experiment. To avoid scattering by the edges, the laser beam is focused on a narrow strip $z_0 \pm \delta z$ in the 7 mm side of the sample, where $\delta z \approx 50 \mu$m and $z_0 \approx 100 \mu$m, counting from the top surface.}
\end{center}
\end{figure}

A very different geometry is used in this work: the luminescence is excited at an edge side of the wafer and observed from the broadside, Fig. \ref{geom}.  The luminescence intensity measured as a function of the distance $x$ from the edge is directly proportional to the hole distribution $p(x)$ in the sample. In a moderately doped sample, we observe a power-law distribution, providing an unambiguous manifestation of the L\'evy  flight of holes. In samples of heavier doping, $p(x)$ evolves toward a distribution characteristic of truncated Levy flight. 


We have studied three $n$-InP samples \cite{NIKKO} of thickness $d$=350 $\mu$m and doping $N_d=$ 0.3, ~2 and~ 6$\times 10^{18}$ cm$^{-3}$ (samples I, II and III, respectively). A 808-nm laser beam was focused (see Fig. \ref{geom}) on the 7-mm edge side of the sample using a cylindrical lens. The excitation flux was uniform along the edge. Excitation photon energy $E =1.53$ eV (above the interband absorption edge at $E_g \approx 1.35$ eV) ensured the initial hole generation in a thin layer near the surface. Luminescence emitted from the broadside was captured by a lens to project the image on a CCD camera. A razor blade was installed near the excitation edge to obscure the luminescence emitted in the direction of the camera from the edge surface itself.

\begin{figure}[b]
\begin{center}
\epsfig{figure=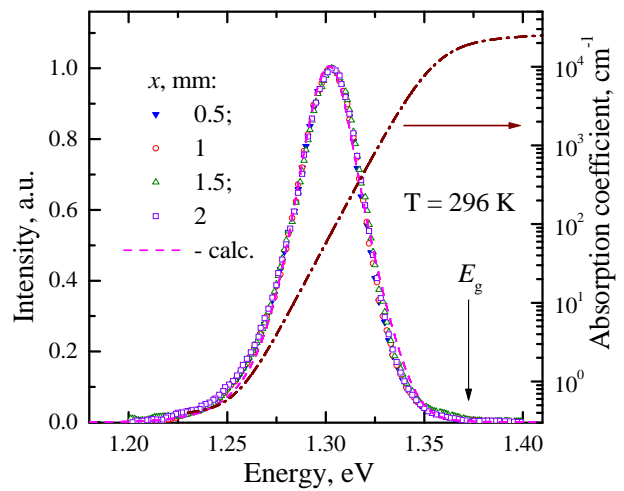,width=8.1cm,height=6.5cm} \caption[]
{\label{spec} Luminescence spectra observed from the top side of sample I at
varying distances from the edge (solid lines). Dashed line shows the calculated spectra assuming an intrinsic van Roosbroek-Shockley emission spectrum, modified by reabsorption of radiation in the sample. Also shown is the absorption spectrum $\alpha (E)$ exhibiting an Urbach tail of $\alpha_i$ below $E_g$ in a wide range of variation. The Urbach-tail dependence is masked by the free-carrier absorption only for $E< 1.26$ eV.  }
\end{center}
\end{figure}

We have also measured the luminescence spectra at different distances $x$ from the excitation. These normalized spectra are shown in Fig. \ref{spec} for sample I. In the range of $x$ from 0.5 to 3 mm, the spectra remain essentially the same and in agreement with the calculated spectrum (dashed line) for a near-uniform hole distribution $p(z) = p(x_0 ,z)$ $\approx {\rm const}$, as expected for any fixed $x_0 \ge d$ (see below).

The absorption spectrum $\alpha(E)$ of sample I is shown in Fig. \ref{spec} by a dash-dotted line. It exhibits an exponential Urbach behavior, extending down to $\alpha \approx 0.5$ cm$^{-1}$. At lower $E$, the Urbach tail, characteristic of interband absorption $\alpha_i$, is masked by the free-carrier absorption, $\alpha(E)=\alpha_i(E)+\alpha_{fc}$. At high $E$, the  Urbach exponent saturates above $E_g$. With the higher doping, the absorption edge is shifted to higher energies due to the Moss-Burstein effect. It is notable that the observed luminescence spectrum lies fully in the Urbach tail region of the absorption spectrum  --- in contrast with the intrinsic emission spectrum described by the quasi-equilibrium van Roosbroek-Shockley \cite{VRS} relation (VRS),
\be
S_{VRS}(E) \sim \alpha_i E^3 e^{-E/kT}~,
\label{VRS}
\ee
which has a maximum above $E_g$.

\begin{figure}[t]
\begin{center}
\epsfig{figure=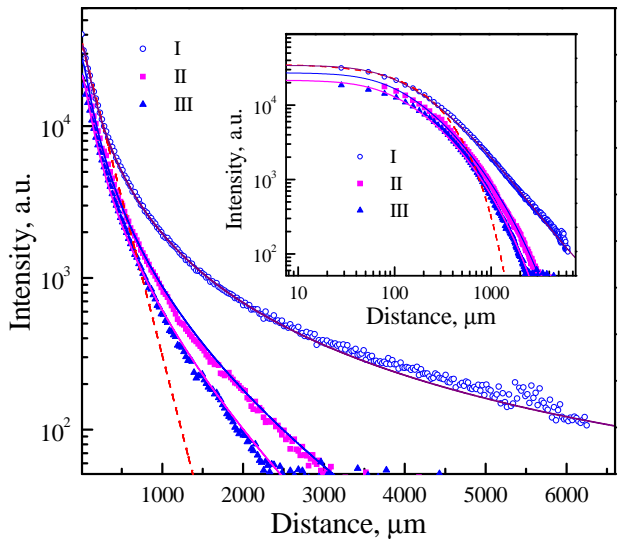,width=8.2cm,height=7.2cm} \caption[]
{\label{dist} Distribution of the luminescence intensity $I(x)$ and hole concentration $p(x) \propto I(x)$ for three differently doped samples (dots). Results of calculations using optical parameters of the samples are shown by solid lines. For comparison also shown (by a dashed line) an exponential distribution with enlarged diffusion length $l=210$ $\mu$m.  For the low-doped sample I, the distribution is a power law, with the index $\gamma$ readily obtained from the slope of the log-log scaled graph in the inset.  For samples II and III with higher $N_d$, the distribution is truncated by free-carrier absorption.  }
\end{center}
\end{figure}

The luminescence intensity distributions $I(x)$ were obtained by scanning the CCD image near its center along a line parallel to $x$. To reduce random fluctuations, the distributions were averaged over 20 scans. Additionally, the CCD camera was shifted both in the $y$ and $x$ directions, with the corresponding scans averaged again to reduce irregularities in the response of different pixels. The observed $I(x)$ was strictly proportional to the excitation laser power and hence to the hole concentration $p(x)$. To stay within the linear-response range of CCD over the entire range of $x$, we first used a neutral filter to reduce $I(x)$ near the excitation edge and then the razor blade was shifted along $x$ to obscure the brightest parts of the slab just near the edge. Subsequently, the filter was removed to get a measurable signal far from the edge. The residual dark noise of the camera was subtracted. The resulting distributions for all three samples are presented in Fig.  \ref{dist}.

For sample I at distances $x>0.5$ mm, a power law $I(x) \sim 1/x^{1+\gamma}$ is clearly observed. This is most easily seen on the log-log scale in the inset of Fig.~ \ref{dist}. The best fit for $\gamma =0.7 \pm 0.1$ agrees with the index $\gamma$ obtained earlier \cite{Semyon1} for ${\cal P} (x)$. The power law is in clear contrast to an exponential decay $I(x) \sim \exp (-x/l)$  expected for a (photon-assisted) diffusive spread of holes, even accounting for any enhancement of the diffusion length $l$ by recycling (a diffusive curve for $l$=210 $\mu$m is shown in Fig.~ \ref{dist} by the dashed line).

For samples II and III with higher doping, the heavy tails are also clearly seen. However, the power-law distribution is truncated at large distances. This effect correlates with the increasing $\alpha_{fc}(N_d)$.  

The nature of the emission spectra is discussed in detail in the review \cite{Luryi}. The room-temperature intrinsic emission spectrum in $n$-InP is well described by Eq. (\ref{VRS}). The observed spectrum $S(E)$ is modified (filtered) by reabsorption on its way out of the sample, $S(E)=S_{VRS}\times F(E)$, where $F(E)$ is the radiation filtering function $F(E)=F_1(E)~T(E)$ which depends on the hole distribution $p(z)$ across the wafer and is affected by reflections from the sample surfaces. The one-pass filtering function $F_1(E)$ is given by \cite{Semyon1,Luryi}
\be
F_1(E)=\int_0^d p(z)\exp[-\alpha(E)z]dz~.
\label{Filt}
\ee
The factor $T(E)=[1-R\exp(-\alpha(E) d)]^{-1}$ accounts for the multiple surface reflections. Note that due to the high index contrast the outgoing radiation propagates close to the normal direction to the surface and the surface reflection coefficient $R(E)\approx 0.33$ varies little in the narrow energy range of the emission line.  Multiple reflections are noticeable in the red wing of the spectrum.  

One can expect that for $x>d$ the distribution $p(z)$ is uniform across the sample, except for small regions near the surfaces, where it is distorted by surface recombination. This assumption is confirmed by solving the one-dimensional diffusion equation with a recycling term [see Eq. (\ref{diff_reequ}) below] for $p(z)$ with uniform generation at a given $x$. The calculated emission spectrum is fully determined by parameters of the absorption spectrum. Results of the calculations, shown in Fig. \ref{spec} by a dashed line, agree with experiment near perfectly, without adjustable parameters. Position of the emission line maximum can be calculated analytically by substituting $p(z)=~$const into Eq. (\ref{Filt}). Near the center of the emission line the absorption coefficient is described by a simple exponent $\alpha=\alpha_0\exp[(E-E_g)/\Delta]$ with the parameters, $\alpha_0 = 1.1\times 10^4$ cm$^{-1}$, $E_g = 1.354$ eV, $\Delta=9.4$ meV, known from optical studies \cite{AOS,Urb}. The emission line peak position $E_{\rm max}$ is then found from the equation $d S(E)/dE=0$:
\be
E_{\rm max}=E_g-\Delta\ln(\alpha_0d/s)~,
\ee
where $s$ is a non-zero solution of transcendental equation $(kT/\Delta)s=[\exp(s)-1]$.  This gives $E_{\rm max}=1.303$ eV, in agreement with experiment. This agreement further supports our assumption of the shape (\ref{VRS}) for the intrinsic spectrum and confirms the proportionality $p(x) \propto I(x)$.

Next we discuss the hole distribution $p(x)$ along the sample. Since the excitation is restricted to a narrow region near the edge surface, and all surfaces are highly reflective (with a narrow radiation escape cone),  this distribution can be found from the one-dimensional Biberman-Holstein stationary transport equation \cite{Ivanov,Bib,Holstein,Luryi}
\be
  - D \frac{ \partial^2
 p(x)}{\partial x^2}+\frac{ p(x)}{\tau}
= G(x)+\frac{\eta}{\tau}  \int
_{-\infty}^\infty   p(x^\prime) {\cal P}( |x-x^\prime|)dz^\prime  \label{diff_reequ}
\ee
where $D$ is the ordinary hole diffusivity and $\tau^{-1}$ is the total recombination rate. The last term in the right-hand side accounts for the photon recycling. The reflection from the edge face at $x=0$ is included by assuming a symmetric distribution  $p(x)=p(-x)$ and extending the integration to $-\infty$ (which supplies an image source for every radiative recombination event in the sample). Here ${\cal P}(|x-x^\prime|)$ is the probability, averaged over the plane $x=$ const, of a hole to generate another hole at a distance $|x-x^\prime|$ by the emission-reabsorption process (see \cite{Luryi}):
\be
{\cal P}(|x|)=\frac{1}{2} \int_0^\infty {\cal N} (E) \alpha_i(E) {\rm Ei}(1, \alpha(E) |x|)  dE \label{Kernel}~,
\ee
where 
${\rm Ei}(1, x)=\int dt~t^{-1}\exp(-xt)$. The integrand in Eq. (\ref{Kernel}) is a product of probabilities for (i) emission of a photon at energy $E$, described by a normalized spectral function ${\cal N}(E)\propto E^{-1}S_{VRS}(E)$, (ii) propagation of this photon from a point at distance  $x$ to another point at distance $x^\prime$, described by the factor ${\rm Ei}(1, \alpha(E) |x|)$, and (iii) interband absorption of this photon, represented by the factor $\alpha_i(E)$. Hence, neglecting the very small effects of hole displacement in their thermal motion, the function ${\cal P}(|x-x^\prime|)$  is the single-step length distribution for holes. The probability ${\cal P}(|x|)$ given by Eq. (\ref{Kernel}) is again fully determined by the absorption spectrum and can be calculated numerically \cite{Semyon1,Luryi}. For all  $x \ge 1 \mu$m, it obeys the power law (\ref{StepDistrTrunc}) with an exponential factor allowing for free-carrier absorption. The index of the distribution is fully determined by the Urbach-tail part of the spectrum and is given by $\gamma = 1- \Delta/kT$.    

Solution of Eq. (\ref{diff_reequ})  can be obtained by a Fourier transformation. For $G=G_0\delta(x)$ it is of the form
\be
p(x)=\frac{G_0\tau}{\pi}\int_0^\infty \frac{\cos(kx)}{1+l^2k^2-\eta F(k)}dk~.
\label{SolAD}
\ee
Here $l=\sqrt{D\tau}$ is the ordinary hole diffusion length, and $F(k)$ is the Fourier-transform of ${\cal P}(|x|)$. For $x\gg l$ the distribution $p(x)$ is fully determined by $F(k)$ (and hence in turn by the absorption spectrum). The only additional parameter  $\eta$ has been already determined for each sample --- from the time-resolved luminescence kinetics experiments \cite{Semyon2} and, independently, by analyzing the ratio of transmission-to-reflection luminescence spectra \cite{Semyon1}.

An alternative to solving Eq. (\ref{diff_reequ}) is to do Monte Carlo modeling of the hole distribution using the single-step probability (\ref{Kernel})  \cite{Semyon1,Luryi}.
The result coincides with (\ref{SolAD}) at distances $x<500$ $\mu$m. However, at considerably larger distances of interest to us here, the Monte Carlo approach becomes noisy and therefore less reliable \cite{Clauset}.

Results of the numerical calculations of $p(z)$ using (\ref{SolAD}) for all samples are shown by the solid lines in Fig. \ref{dist} demonstrating an excellent agreement with the experimental data, except for a slight discrepancy for the heaviest doped sample III (see below). 

Main features of the distribution are revealed by an analytic approximation that can be derived directly from Eq. (\ref{diff_reequ}). At large distances, due to the heavy tail of ${\cal P}(|x|)$, one can solve Eq. (\ref{diff_reequ}) by sequential iterations --- making in all terms of the resultant series the ``longest step approximation'' \cite{Ivanov} (i.e. choosing one of the steps equal the total distance $x$). This gives
\be
p(x)=\frac{\Phi{\cal P}(x)}{ [1+\Phi {\cal P}_e(x)]^2}~,
\label{interp}
\ee
where ${\cal P}_e(x)= \int_x^\infty{\cal P}(x')dx'$ is the probability of escape beyond $x$ in one step. Equation (\ref{interp}) provides a good approximation to the exact solution in the entire range of $x$. It works with or without truncation of the L\'evy flight by free-carrier absorption. Similarly to the ``stable distribution'' \cite{Levy}, distribution (\ref{interp}) asymptotically reproduces the one-step probability  (enhanced by the recycling factor $\Phi$) and is modified at smaller distances. Since $\Phi{\cal P}_e(x)$ is the escape probability in $\Phi$ attempts, condition $\Phi {\cal  P}_e(x_f)=1$  gives the distance $x_f$ to the ``front'' of $p(x)$ --- beyond which the holes appear predominantly in one step. The front $x_f$ is clearly seen in the inset of Fig. \ref{dist}, as the point of maximum curvature on the log-log plots for all samples. The observed  $x_f \approx 200$ $\mu$m manifests the large values of $\Phi$ in these samples. For $x \ll x_f$, Eq. (\ref{interp}) correctly predicts the short-distance asymptote \cite{Luryi}, $p(x) \sim 1/x^{1-\gamma}$ but the corresponding data are obscured in our experiment by deviations from the one-dimensional geometry \cite {NoteZo}. For sample III, the effective front distance $x_f$ is somewhat larger than could be expected from earlier experiments.

In conclusion, we have studied the stationary hole distribution $p(x)$ over distances $x$ of several mm from initial photoexcitation.  We observe a heavy-tailed decline of the luminescence intensity with no change in the spectral shape. For a low-doped sample, we find a power-law distribution $p(x)$ precisely accounted for by the L\'evy-flight transport of holes mediated by photon recycling.  In heavier-doped samples, the power law is truncated by free-carrier absorption, which manifest itself only  at distances $x\sim \alpha_{fc}^{-1}$ corresponding to the free-carrier absorption. The anomalous transport should be important in all semiconductor crystals with high radiative efficiency and may have practical implications in optoelectronic devices \cite{Luryi}. It is of interest to study enhancement of this effect at lower temperatures. 

This work was supported by the Domestic Nuclear Detection Office, by the Defense Threat Reduction Agency (basic research program), and by the Center for Advanced Sensor Technology at Stony Brook. 


\begin{thebibliography}{99}
\bibitem{NoteZo} Geometry of the photon transport deviates from one-dimensional at short distances ($x \sim d/2$), since the excitation beam is focused and nonuniform in $z$ direction. This effect is taken into account empirically by the replacement $x\rightarrow \sqrt{x^2+z_0^2}$ with $z_0 \approx 60$ $\mu$m as a fitting parameter. This does not affect the results at $x \ge d$.
\end{thebibliography}
\end{document}